# On a Pregeometric Origin for Spacetime Dimensionality and Metric Structure

W.M. Stuckey
Dept of Physics
Elizabethtown College
Elizabethtown, PA  17022
stuckeym@etown.edu
(717) 361-1436 (office)
(717) 361-1176 (fax)

**Abstract**
Motivated by Wheeler's 'bottom up' pregeometry, we introduce a pregeometric approach that does not assume Wheeler's probability amplitudes for establishing spacetime neighborhoods. Rather, a non-trivial metric is produced via the concept of a *uniformity base*, which is generated with discrete topological groups over some arbitrary fundamental, denumerable set. We show how the concept of *entourage multiplication* for the elements of our uniformity base mirrors the underlying group structure. This fact is then exploited to create entourage sequences of maximal length, whence a fine metric structure. The resulting metric structure is, for certain group structures, consistent with E4-embeddable graphs. Examples over $Z_2 \times Z_4$, $D_4$, $Z_6$, $D_3$, $Z_8$, and $Z_5$ are provided. Euclidean embeddability over $Z_7$ and $Q_8$ is discussed. Unlike the statistical approaches typical of graph theory, this method generates dimensionality over low-order sets. Possible applications to the pregeometric modeling of quantum stochasticity and non-locality/non-separability, wave function collapse, and the M4 structure of spacetime are provided in the context of $Z_2 \times Z_2 \times Z_4$.

# 1. INTRODUCTION

According to Wheeler's pregeometry, properties of the spacetime manifold, such as metric, continuity, dimensionality, topology, locality, symmetry, and causality might evolve mathematically from modeling underlying classical spacetime dynamics (Gibbs, 1998). This approach has been labeled "bottom up" by Requardt & Roy (2001) in contrast to the "top down" approach originally proposed by Sakharov (1967) and developed more recently by Akama & Oda (1992), and Terazawa (1992). In top down pregeometry, particle physics is assumed fundamental to the spacetime geometry while Wheeler assumed "the order of progress may not be physics $\rightarrow$ pregeometry, but pregeometry $\rightarrow$ physics" (Misner, Thorne & Wheeler, 1973).

In an early attempt to derive dimensionality, Wheeler assigned probability amplitudes to the members of a Borel set[1] to stochastically establish spacetime adjacency (Wheeler, 1964). He abandoned this idea, in part, because "too much geometric structure is presupposed to lead to a believable theory of geometric structure" (Wheeler, 1980). In particular, he considered the manner in which probability amplitudes were assigned, and a metric introduced, to be *ad hoc*. However, recent models by Nagels (1985), Antonsen (1994), and Nowotny & Requardt (1998) employing graph theory have, arguably, surmounted these objections.

In these approaches, assumptions concerning the stochastic analyses are minimal and natural. Nagels (1985), for example, assumes only that the order of the underlying set is large and the probability of adjacency is small. And, the notion of length per graph theory is virtually innate.

---

1. "Loosely speaking, a Borel set is a collection of points which have not yet been assembled into a manifold of any particular dimensionality. Whether they are put together into lines, or surfaces, or volumes, or manifolds of higher dimensionality, or into odd combinations of such objects of varied dimensionality, is a matter of choice." (Wheeler, 1964)

Thus, Wheeler's pregeometric progeny have managed to obtain dimensionality, in various forms, over otherwise structureless sets.

Of course, one might contend that the connections between set members (bonds, links, edges) provide excessive *a priori* structure. However, connections are as fundamental to graph theory as the set itself. And we note, insofar as pregeometry is a search for a fundamental language of physics, the graph theoretical approach resonates with Heylighen's suggestion that a fundamental language of physics might be couched in "processes" (Heylighen, 1990). His "events and arrows" parallel the "points and links" of the graph theoretical approaches *supra*. Antonsen (1994), for example, refers to his links as "interactions" and defines "corresponding 'second quantization' operators b, $b^+$." And, Requardt (2000) refers to "dynamical bonds which transfer the elementary pieces of information among the nodes." It bodes well for the graph theoretical approach that a self-consistency relationship between spacetime geometry and physical processes – a pregeometric version of Einstein's equations – is already implied.

Given the promising nature of graph theory for bottom up pregeometry, we introduce a complementary approach. Whereby in graph theory the connection is fundamental to the metric and large-order sets are employed, our approach uses small-order sets and connections are inferred from the metric. This type of bottom up pregeometry may have bearing on the modeling of quantum phenomena such as non-locality, where one might expect the structure of microscopic spacetime neighborhoods differs from that of macroscopic spacetime neighborhoods (Stuckey, 1999). To contrast the statistical approach, we introduce uniformity bases induced by discrete topological groups. The use of metrizable uniformities in bottom up pregeometry has already proven effective in producing spacetime dimensionality (Bergliaffa *et al.*, 1998). In our approach, all subsets are open – the discrete topology, as with graph theory (Nowotny &

Requardt, 1999) – and any group structure permissible. We begin by describing our construct of a uniformity base over a discrete topological group.

## 2. THE UNIFORMITY BASE, ENTOURAGE SEQUENCE AND METRIC

We refer to the underlying set X of order N as a Borel set, since it is denumerable, finite and structureless per Wheeler's description, and its members needn't be zero-dimensional objects. The category of uniform spaces subsumes topological groups (Miyata, 1993) and a topological group may be created over any group with the discrete topology. Therefore, in the spirit of avoiding arbitrary assumptions, we assume all subsets of X are open and that we are free to consider any group structure G of order N.

To review briefly, a *uniformity over a set X* (or *uniform space*) is constructed from collections of pairs $(x, y) \in X \times X$. Each collection of pairs is called an *entourage*. Entourages A and B may be multiplied in an obvious fashion, i.e., AB is the collection of pairs $(x, z)$ such that $(x, w)$ is in A and $(w, z)$ is in B. The collection of entourages must satisfy the following properties in order that they constitute a uniform space:

1. The intersection of all entourages is precisely $\Delta \equiv \{(x, x) \mid x \in X\} \; \forall \; x \in X$. The entourage $\Delta$ is called the *diagonal*.
2. If A is an entourage, $A^{-1} \equiv \{(w, z) \mid (z, w) \in A\}$ is an entourage.
3. If A is an entourage, there exists an entourage B such that $BB \subset A$.
4. The intersection of two entourages is an entourage, and the superset of an entourage is an entourage.

We construct a *uniformity base* $U_B$ for the uniformity U via neighborhoods of the identity e of G per Geroch (1985). The entourage $A_\alpha$ of U is $\{(x, y) \in X \times X \mid xy^{-1} \in \alpha\}$ where $\alpha$ is a neighborhood of e in the topology over X. With X denumerable of order N, $\{(w, y) \in X \times X \mid w \neq y\}$ is partitioned equally into the entourages $A_x$ ($x \in X$ such that $x \neq e$) for the N – 1, order-two neighborhoods of e, i.e., $A_x$ is generated by {e, x}. The entourages $A_x$ and $\Delta$ constitute our uniformity base $U_B$ for U.

In order to produce a metric from this uniformity base, we borrow from a proof of the following theorem by Engelking (1989).

"For every sequence $V_0$, $V_1$, ... of members of a uniformity on a set S, where

$$V_0 = X \times X \text{ and } (V_{i+1})^3 \subset V_i \text{ for } i = 1,2,...,$$

there exists a pseudometric $\rho$ on the set S such that for every $i \geq 1$

$$\{(x, y) \mid \rho(x, y) < (½)^i\} \subset V_i \subset \{(x, y) \mid \rho(x, y) \leq (½)^i\}."$$

To find $\rho(x, y)$, consider all sequences of elements of S beginning with x and ending with y. For each adjacent pair $(x_n, x_{n+1})$ in any given sequence, find the smallest member of $\{V_i\}$ containing that pair. [The smallest $V_i$ will have the largest i, since $(V_{i+1})^3 \subset V_i$.] Suppose $V_m$ is that smallest member and let the 'artificial' distance between $x_n$ and $x_{n+1}$ be $(½)^m$. Summing for all adjacent pairs in a given sequence yields an 'artificial' distance between x and y for that particular sequence. According to the theorem, $\rho(x, y)$ is the greatest lowest bound obtained via the sequences.

Since our set X is denumerable and finite, this greatest lower bound will be non-zero and the result will be a metric. In order to obtain a metric with maximal resolution (fine metric) per this formalism, we need an entourage sequence $V_0$, $V_1$, ... of maximal length. In order to produce such a sequence from the members of $U_B$, we note the following two properties.

First, not all the elements of our $U_B$ will be symmetric as required by Engelking's proof. In fact, $A_x$ is symmetric for all $x \in X$ such that $x = x^{-1}$. This, since for $(y, z) \in A_x$ such that $y \neq z$, $yz^{-1} = x$ and therefore, $zy^{-1} = x^{-1} = x \Rightarrow (z, y) \in A_x$. For the base members $A_x$ and $A_y$ such that $x = y^{-1}$, we have $A_x^{-1} = A_y$. This, since for $(w, z) \in A_x$ such that $w \neq z$, $wz^{-1} = x$ and therefore, $zw^{-1} = x^{-1} = y \Rightarrow (z, w) \in A_y$. Accordingly, for $x = y^{-1}$ we must have $A_x \cup A_y$ appear in any $V_i$ to maintain the symmetry required by Engelking's theorem.

Second, we show that entourage multiplication of the $A_x \in U_B$ mirrors the underlying group structure. With $\Delta$ a subset of any entourage (uniquely and axiomatically), we have in general for entourages A and B that $A \subset AB$ and $B \subset AB$. Now consider $\{(x, y), (y, z) \mid (x, y) \in A_s$ and $(y, z) \in A_w$ with $x \neq y$ and $y \neq z\}$. In addition to $\Delta$, these account exhaustively for the elements of $A_s$ and $A_w$. For any such pair $(x, y)$ and $(y, z)$, $(x, z) \in A_s A_w$ by definition and $(x, z) \in A_{sw}$, since $sw = (xy^{-1})(yz^{-1}) = xz^{-1}$. The N pairs $(x, z)$ with $\Delta$ account exhaustively for the elements of $A_{sw}$ and, excepting the impact of $\Delta$ on $A_s A_w$, the N pairs $(x, z)$ account exhaustively for the elements of $A_s A_w$. Again, the impact of $\Delta$ on $A_s A_w$ is to render $A_s \subset A_s A_w$ and $A_w \subset A_s A_w$. Therefore, $A_s A_w = A_s \cup A_w \cup A_{sw}$. Accordingly, we may exploit subgroup structure to produce an entourage sequence of maximal length. Nested subgroup structures are particularly useful, since we require $(V_{i+1})^3 \subset V_i$. As an example of this approach, and to illustrate how it can generate dimensionality, we obtain a fine metric over $Z_2 \times Z_4$.

## 3. METRIC AND DIMENSIONALITY OVER $Z_2$ x $Z_4$

Let $X = \{a, b, c, e, A, B, C, D\}$ with the lower case elements closed under the algebra of the $Z_4$ (sub)group where $b = b^{-1}$ and e is the identity. With 0 and 1 the typical elements of $Z_2$, the upper case elements of X can be identified as $A = (1, a)$, $B = (1, b)$, $C = (1, c)$, and $D = (1, e)$. We have a nested subgroup structure of $Z_2 \subset Z_4 \subset Z_2 \times Z_4$, so we choose the following entourage sequence via $U_B$ :

$$V_3 = A_b$$

$$V_2 = A_b \cup A_a \cup A_c$$

$$V_1 = A_b \cup A_a \cup A_c \cup A_D$$

$$V_0 = A_b \cup A_a \cup A_c \cup A_D \cup A_A \cup A_B \cup A_C$$

which yields the following hierarchy for 'artificial' distances between pairs:

{(a,c), (b,e), (A,C), (B,D)} $\Rightarrow$ 1/8            [$xy^{-1} = b$]

{(a,e), (b,a), (c,b), (e,c), (A,D), (B,A), (C,B), (D,C)} $\Rightarrow$ 1/4      [$xy^{-1}$ = a or c]

{(a,A), (b,B), (c,C), (e,D)} $\Rightarrow$ 1/2               [$xy^{-1} = D$]

{(a,D), (b,A), (c,B), (e,C), (A,e), (B,a), (C,b), (D,c)} $\Rightarrow$ 1       [$xy^{-1} = B$]

{(a,C), (b,D), (c,A), (e,B), (A,c)} $\Rightarrow$ 1              [$xy^{-1}$ = A or C].

Therefore, we have the following metric structure (after a convenient renormalization):

| g(x,y) = 1 | g(x,y) = 2 | g(x,y) = 4 | g(x,y) = 5 | g(x,y) = 6 |
|---|---|---|---|---|
| (a,c) | (a,e) | (a,A) | (a,C) | (a,B) |
| (b,e) | (b,a) | (b,B) | (b,D) | (b,C) |
| (A,C) | (c,b) | (c,C) | (c,A) | (c,D) |
| (B,D) | (e,c) | (e,D) | (e,B) | (e,A) |
|  | (A,D) |  |  | (C,e) |
|  | (B,A) |  |  | (D,a) |
|  | (C,B) |  |  | (A,b) |
|  | (D,C) |  |  | (B,c) |

This is consistent with two, triangular polyhedra occupying E3's (figure 1) that are embedded in E4. In this embedding, the distances 1, 2, and 4 are understood as straight line (direct) distances while the distances 5 and 6 are obtained along indirect paths. For example, $g(e,A) = 6$ obtains along the path $e \to a \to A$, since $g(e,a) = 2$ and $g(a,A) = 4$. [The path-dependent interpretation of g obtains with any embedding, tacitly if not explicitly.] Thus, the group structure $Z_2 \times Z_4$ induces an E4-embedded graph. In a similar manner, $D_4$, $Z_6$, $D_3$, $Z_8$, and $Z_5$ generate dimensionality via Euclidean-embedded graphs.

## 4. OTHER EXAMPLES

We first consider the fourth dihedral group ($D_4$ – symmetry group over the square) for reasons immediately apparent. A maximal entourage sequence for $D_4$ contains four entourages. $V_3$ and $V_2$ are generated by the rotations – the $Z_4$ subgroup – in exactly the same manner as $Z_2 \times Z_4$. Adding any member of $U_B$ based on a reflection to $V_2$ yields $V_1$, since the reflections are their own inverses. And, of course, $V_0 = X \times X$. Perhaps not surprisingly, given the nested $Z_2 \subset Z_4 \subset D_4$ structure, the embedded graph is precisely that generated by $Z_2 \times Z_4$ (figure 1). Choosing to exploit the $Z_2 \subset Z_2 \times Z_2 \subset D_4$ nested structure yields the same embedded structure with permuted nodes. [The possible use of such ambiguity in the modeling of quantum stochasticity is discussed in section 5.] That two distinct group structures generate the same embedded graph also obtains with $Z_6$ and $D_3$.

We choose the typical representation of $Z_6$, addition mod 6 over $\{0,1,2,3,4,5\}$. Then a maximal entourage sequence from $U_B$ is:

$$V_2 = A_2 \cup A_4 \supset \{(2,0), (4,0), (2,4), (1,5), (1,3), (3,5)\}$$

$$V_1 = A_2 \cup A_4 \cup A_3 \supset \{(2,5), (4,1), (0,3)\}$$

$$V_0 = A_1 \cup A_2 \cup A_3 \cup A_4 \cup A_5 = X \times X.$$

Since $V_2$ is constructed via the $Z_3$ subgroup, 0, 2 and 4 are paired in $V_2$ as are their complements 1, 3 and 5. Therefore, the fundamental Euclidean-embeddable structures are equilateral triangles (E2-embeddable). These can be connected in E3 (figure 2), since $V_1$ can be constructed from $V_2$ by adding one additional element of $U_B$, i.e., $A_3$. [That $Z_6$ is isomorphic to $Z_2 \times Z_3$ is manifested in this structure.] We could also choose to exploit the $Z_2$ subgroup structure, i.e., let $V_2 = A_3$. In that case we obtain the same E3 structure with different scaling (figure 3). $D_3$ (symmetry group over the equilateral triangle) generates the same E3 structure as found in figure 2, since it also contains the $Z_3$ subgroup (rotations) upon which $V_2$ is based, and adding any member of $U_B$ based on a reflection to $V_2$ yields $V_1$.

Another group that generates three Euclidean dimensions is $Z_8$. We choose the typical representation of $Z_8$, addition mod 8 over $\{0,1,2,3,4,5,6,7\}$. We've a $Z_2$ subgroup, so the first member of our entourage sequence is $A_4$. If we add $A_2$ and $A_6$ (completing the $Z_4$ subgroup) to construct the next member, our first two entourages would generate two of the E3-embedded graphs of figure 1. However, none of the remaining members of $Z_8$ is its own inverse, so the next simplest member of an entourage sequence will generate too many edges to complete the connection of the E3 polyhedra in E4. Thus, we construct the second member of the sequence by adding $A_3$ and $A_5$ to the first entourage so our sequence is:

$V_2 = A_4 \supset \{(4,0), (1,5), (2,6), (3,7)\}$

$V_1 = A_4 \cup A_3 \cup A_5 \supset \{(0,5), (0,3), (1,6), (1,4), (2,7), (2,5), (4,7), (3,6)\}$

$V_0 = X \times X$.

This entourage sequence does generate a metric whence an E3-embeddable graph (figure 4).

The problem with finding an alternative to the connection of the E3 polyhedra in E4 also obtains with the quaternion group, $Q_8$. Therein resides a $Z_2$ subgroup, but the addition of any other element and its inverse produces a $Z_4$ subgroup (unlike $Z_8$). Thus, there is no Euclidean embedding induced by $Q_8$ per this program.

Hitherto, our examples have yielded three or four Euclidean dimensions. Our last example, $Z_5$, generates but two dimensions. [Note: $Z_N$ is the only group structure of order N when N is prime.] We choose the typical representation of $Z_5$, addition mod 5 over $\{0,1,2,3,4\}$. As a group of prime order there are no subgroups, so we've but two members in our entourage sequence, e.g.,

$V_1 = A_1 \cup A_4 \supset \{(1,0), (0,4), (2,1), (3,2), (4,3)\}$

$V_0 = A_1 \cup A_2 \cup A_3 \cup A_4 = X \times X$.

While there isn't much of a resulting metric structure (all distances 1 or 2), we do obtain an E2-embeddable structure, i.e., the pentagon (figure 5). From this example it is apparent that $Z_7$ will not generate an E4-embeddable structure, encountering precisely the same problems as $Q_8$.

## 5. SPECULATION ON A NEXUS TO PHYSICS

Pregeometry should make correspondence with modern physics, just as modern physics makes correspondence with classical physics. While our pregeometric formalism is as yet inchoate, there are intimations as to how it might make correspondence with quantum stochasticity and non-locality/non-separability, collapse of the wave function, and the relativity of simultaneity – stalwart concepts of modern physics.

Consider $Z_2 \times Z_2 \times Z_4$ and the entangled quantum state

$$|\psi\rangle = \frac{|\uparrow\downarrow\rangle \pm |\downarrow\uparrow\rangle}{\sqrt{2}}.$$

Our group contains the nested subgroup structure $Z_2 \subset Z_4 \subset Z_2 \times Z_4 \subset Z_2 \times Z_2 \times Z_4$, so a maximal entourage sequence is:

$V_5 = A_b$

$V_4 = A_b \cup A_a \cup A_c$

$V_3 = A_b \cup A_a \cup A_c \cup A_D$

$V_2 = A_b \cup A_a \cup A_c \cup A_D \cup A_A \cup A_B \cup A_C$

$V_1 = A_b \cup A_a \cup A_c \cup A_D \cup A_A \cup A_B \cup A_C \cup A_\delta$

$V_0 = X \times X$

where $\delta$ is the counterpart to D in the $Z_2$ duplication. Equivalently, we may replace $V_1$ with $A_b \cup A_a \cup A_c \cup A_D \cup A_A \cup A_B \cup A_C \cup A_\beta$, where $\beta$ is the counterpart to B in the $Z_2$ duplication. Either of these entourage sequences produces a fine metric whence two 'E3-polyhedra-embedded-in-E4' are embedded in E5. The choice of D or B in $V_3$ fixes *both* E4 substructures, so the connection between the two E4 subsets might be interrupted as a non-local/entangled, EPR-type[2] relationship in the otherwise global E4 spacetime structure of quantum mechanics (as opposed to an E5 embedding). Alternatively, the choice of δ or β in $V_1$ might be interpreted as fixing the temporal order of the measurements, which would allow the temporal ordering of space-like separated measurements to be ambiguous, as in the relativity of simultaneity per the M4 spacetime of special relativity.

---

2. EPR stands for Einstein-Podolsky-Rosen, who articulated problems stemming from quantum non-locality in a 1935 publication (*Phys. Rev.* **47**, 777).

In addition to originating with the ambiguity of entourage sequences, quantum stochasticity may also surface via incomplete spatio-temporal boundary conditions. In any theory of physics, a spacetime region is described uniquely only after providing specifics concerning its spatio-temporal boundary. Our approach desribes reality in terms of a unified space and time, i.e., the 'block universe'[3] or spacetime of special relativity. Thus, we expect the unique description of a spacetime region will require information about its future boundary. An inability to supply this information may lead to stochastic, rather than deterministic, descriptions of the spacetime region(s) in question.

In the entangled state *supra*, for example, knowing the temporal order of the measurements corresponds to specifying a particular spacetime foliation. This reduces ambiguity in the entourage sequence (fixes β or δ in $V_1$), thus reducing stochasticity. To determine the spatio-temporal description uniquely requires knowledge of a measurement outcome, which fixes B or D in $V_3$. Thus, per this interpretation, the collapse of the quantum mechanical wave function is explicitly a two-step process in contrast to quantum mechanics whereby the choice of a preferred spacetime foliation is tacit. And, wave function collapse is epistemological – not ontological – as one would expect in a 'block universe' *sans* Everett's many-worlds interpretation (DeWitt & Graham, 1973).

---

3. A 'block universe' needn't be constructed dynamically, i.e., per the temporal evolution of a space-like hypersurface. However, it must subsume our dynamical perspective so as to be open to empirical investigation.

## 6. CONCLUSION

We have shown that metric structure and spacetime dimensionality may be induced over a low-order Borel set per a uniformity base generated via discrete topological groups. The dimensionality produced herein is a simple embedding dimension generated from 'connections' between 'nodes'. Thus, should some of these connections ultimately correspond to physical interactions, our embedding dimension may be viewed in a dynamical spirit *a la* Requardt's "connectivity-dimension" (Requardt, 1996).

Our approach differs from graph theory in that connections are inferred from the metric, which obtains as a natural consequence of algebraic structure. Thus a denumerable, finite set and all its possible group structures are the fundamental constituents of this formalism, which is indifferent to domains of discourse and group representations – a pregeometric counterpart to general covariance, if you will. Since this method is employed over low-order sets, it is intended to provide a pregeometric modeling of microscopic spacetime neighborhoods. And in fact, there are intimations of its potential for modeling quantum stochasticity and non-locality/non-separability, wave function collapse, and M4 spacetime structure. Therefore, this formalism counters Monk's contention that sub-Planck scale structures may not provide an ideal basis for pregeometry, since quantum non-locality/non-separability occurs on macroscopic scales (Monk, 1997).

Finally, contrary to Bergliaffa *et al.* (1998), our pregeometric model does not employ trans-temporal objects as fundamental constituents of reality. In this sense, we agree with Gomatam (1999) who argues for a revision of our notion of macroscopic objects in accord with quantum non-separability. In this sense our pregeometry is amenable to the quantum gravity programs of Heller & Sasin (1999) and Demaret *et al.* (1997), all of whom acknowledge the implications of quantum non-locality/non-separability on the modeling of microscopic spacetime neighborhoods. At this stage, it is probably best that researchers are engaged on all fronts.

## ACKNOWLEDGEMENT

This work was supported in part by the NATO Collaborative Linkage Grant PST.CLG.976850.

**Figure 1**
**Z₂ x Z₄ and D₄**

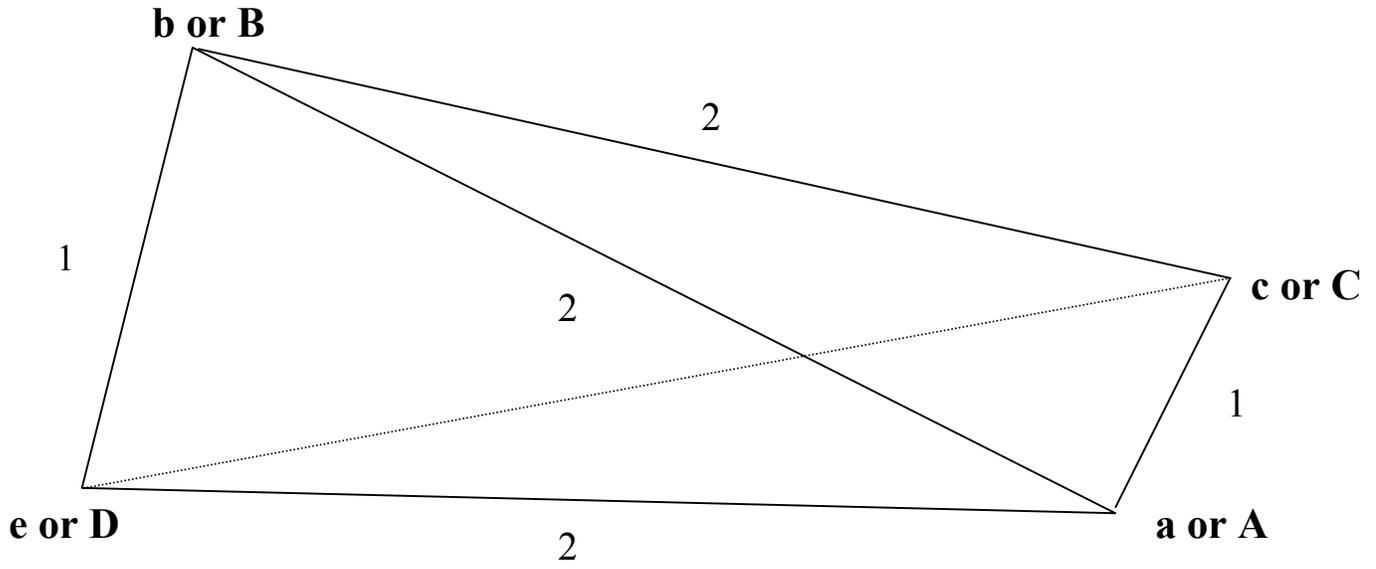

**Figure 2**
**Z_6 and D_3**

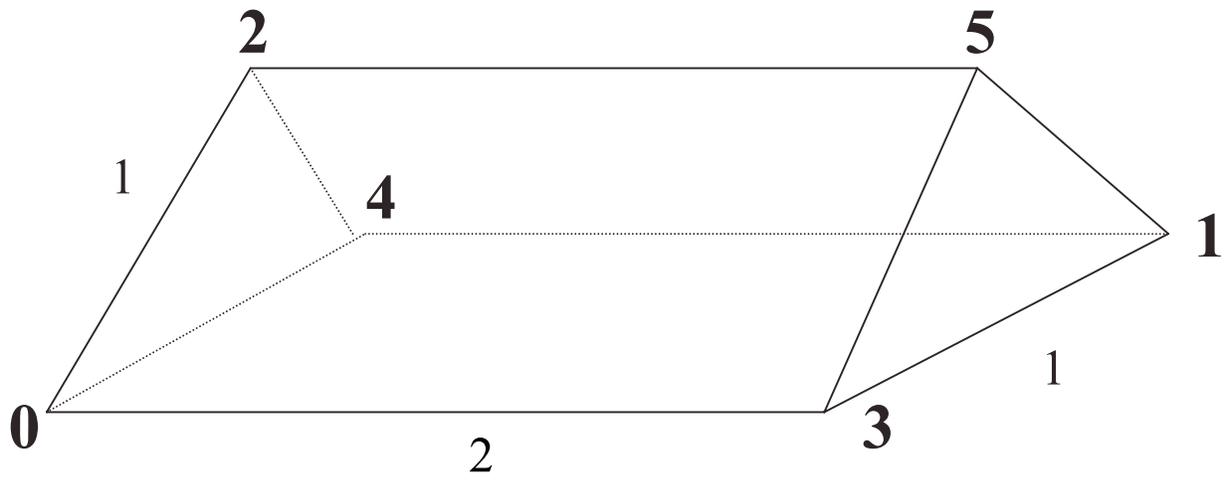

# Figure 3
## $Z_6$

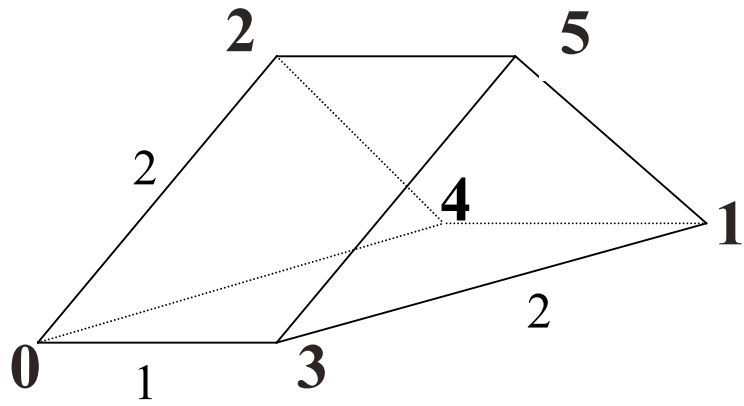

**Figure 4**
**Z₈**

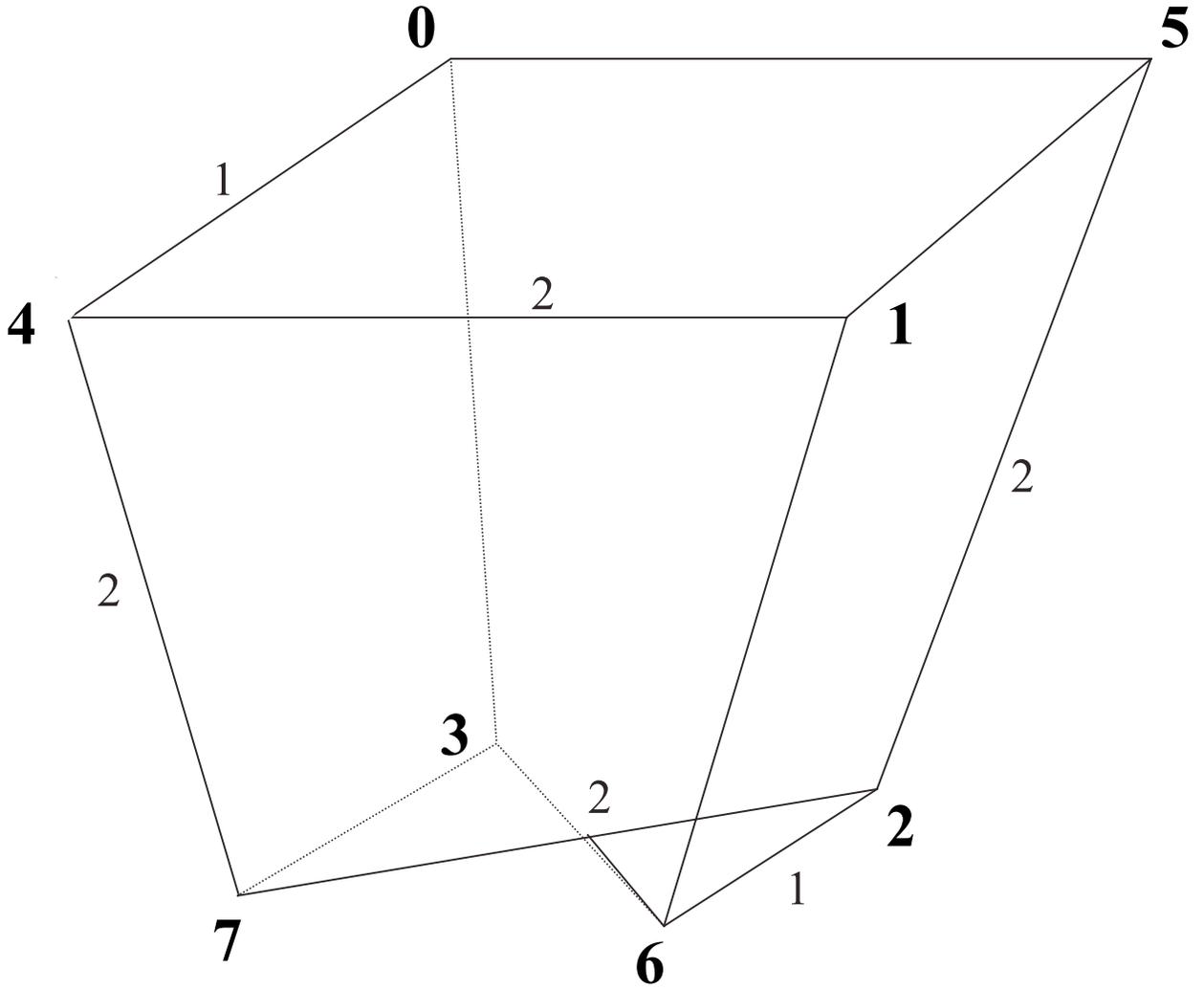

**Figure 5**
**Z₅**

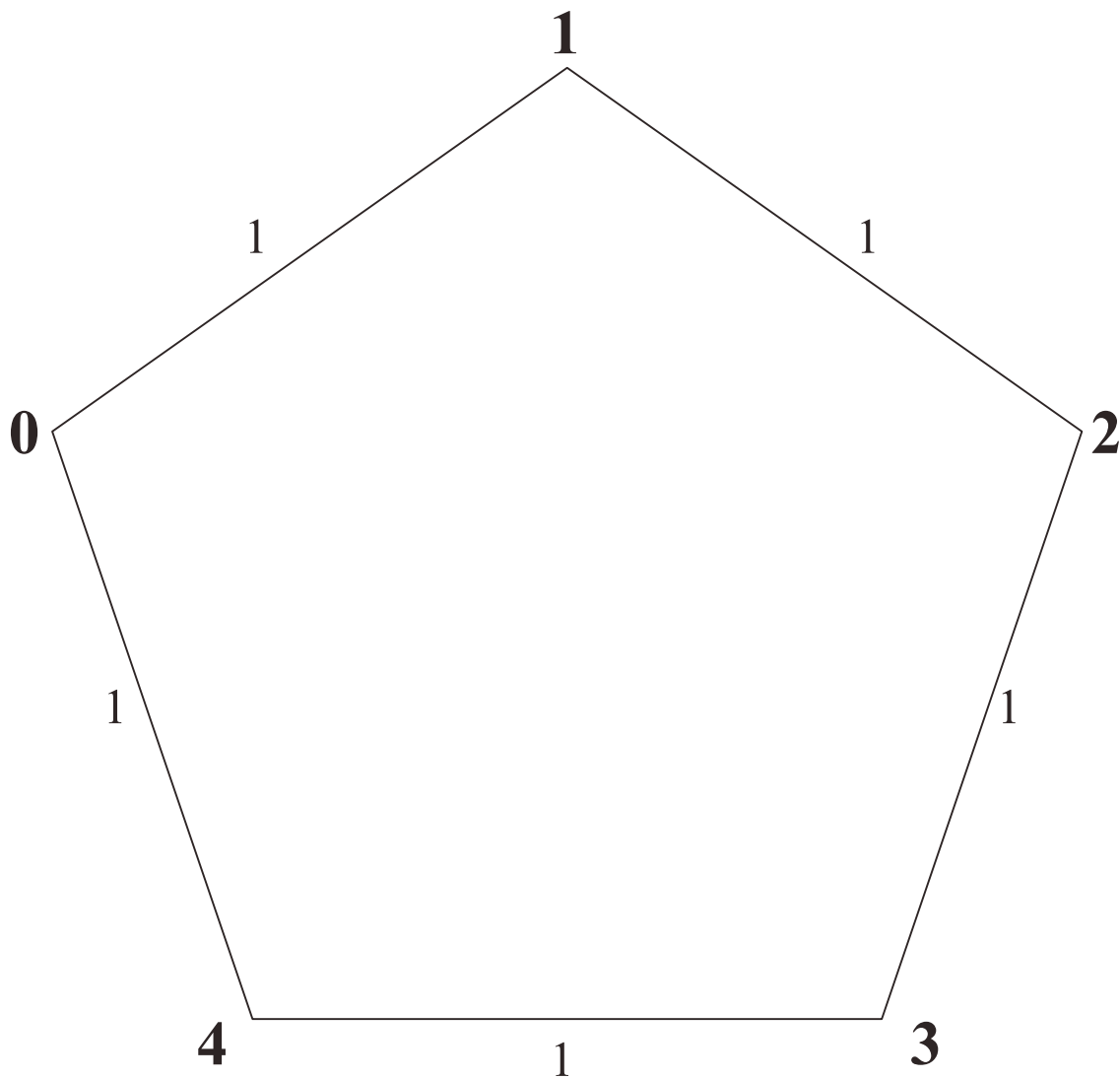